\documentclass[10pt]{article}

\usepackage[english]{babel}
\usepackage[round, comma]{natbib}
\usepackage{appendix} 
\usepackage{color}
\usepackage{multirow}
\usepackage[papersize={8.5in,11in}]{geometry}
 \usepackage[ruled,vlined]{algorithm2e}
\usepackage{caption}
\usepackage{subcaption}
 \usepackage{array}
 \usepackage{multicol}
\usepackage{authblk}
\usepackage{amsmath}
\usepackage{amssymb}
\usepackage{amsthm}
\newtheorem{proposition}{Proposition}

\usepackage{graphicx}
\usepackage{float}
\usepackage{hyperref}
\usepackage{mathtools}
\title{Calibrating the Heston model with deep differential networks}

\author[1,2]{Chen Zhang\thanks{ORCiD: 0000-0003-3481-2739}} 
\author[3]{Giovanni Amici\thanks{ORCiD: 0009-0005-5028-7411. Corresponding author: \href{mailto:giovanni.amici@polito.it}{gamici@ncsu.edu}}}
\author[1]{Marco Morandotti\thanks{ORCiD: 0000-0003-3528-6152}} 

\affil[1]{{\small Department of Mathematical Sciences, Politecnico di Torino 10129, Torino, Italy}}
\affil[2]{{\small Business School, Ningbo University, 315211, Ningbo, China}}
\affil[3]{{\small Department of Industrial and Systems Engineering, North Carolina State University 27606, Raleigh, USA}}

\date{}

\AtBeginDocument{\hypersetup{pdfborder={0 0 1}}}
\usepackage{hyperref}
\usepackage{xcolor}
\usepackage{sectsty}
\definecolor{MyBlue1}{rgb}{0,0,1}
\definecolor{MyBlue2}{rgb}{0.1,0.3,0.8}
\definecolor{MyBlue3}{rgb}{0.3,0.5,1}
\definecolor{MyGreen}{rgb}{0,0.6,0}
\definecolor{MyRed}{rgb}{0.7,0.1,0.1}
\hypersetup{
    linkcolor=black, 
    citecolor=black, 
    linkbordercolor=red, 
    citebordercolor=green, 
    }

\def\GG{\mathcal{G}}
\def\tt{\boldsymbol{\theta}}
\def\xx{\boldsymbol{x}}
\def\yy{\boldsymbol{y}}
\def\WW{\boldsymbol{W}}
\def\bb{\boldsymbol{b}}
\def\pp{\boldsymbol{p}}
\def\dd{\mathrm{d}}
\def\ii{\mathrm{i}}
\def\ee{\mathrm{e}}

\newcommand\norm[1]{\lVert#1\rVert}
\usepackage{comment}
\numberwithin{equation}{section}

\begin{document}

\maketitle
\medskip

\begin{abstract}  
We propose a gradient-based deep learning framework to calibrate the Heston option pricing model
\citep{heston1993closed}.
Our neural network, henceforth deep differential network (DDN), learns both the Heston pricing formula for plain-vanilla options and the partial derivatives with respect to the model parameters. 
%
The price sensitivities estimated by the DDN are not subject to the numerical issues that can be encountered in computing the gradient of the Heston pricing function.
Thus, our network is an excellent pricing engine for fast gradient-based calibrations. 
%
Extensive tests on selected equity markets show that the DDN significantly outperforms non-differential feedforward neural networks in terms of calibration accuracy.
In addition, it dramatically reduces the computational time with respect to global optimizers that do not use gradient information.

\end{abstract}


\noindent \textbf{Acknowledgements:} 
CZ gratefully acknowledges funding from the China Scholarship Council (grant no.~202308330096).
GA is a member of the \emph{Association for Mathematics Applied to Social and Economic Sciences}.
GA and MM are members of the \emph{Gruppo Nazionale per l'Analisi Matematica, la Probabilità e le loro Applicazioni} of the \emph{Istituto Nazionale di alta Matematica}.


\section{Introduction}



Financial derivatives are of core importance for the trading activities of banks and other financial actors.
Their use consists, for example, of hedging positions in primary assets, speculating on the market changes, and designing arbitrage strategies.
Option contracts, in particular, depend on a number of risk factors, such as the underlying asset price, its volatility, and the risk-free interest rate.
As such, in order to estimate the fair value of an option it is crucial to construct models able to accurately describe the dynamics of the most relevant risk factors.
In addition, it is fundamental to design fast calibration techniques that can deal with the frequent changes of the market conditions. 

The famous option pricing model of \citet{black1973pricing} provides a tractable and intuitive framework for the valuation of option contracts.
However, due to its restrictive assumptions (e.g., asset prices following geometric Brownian motions, and constant volatility), the Black-Scholes model fails to accurately capture the market-implied distribution of log-returns.
As such,
over the last decades 
researchers have constructed more sophisticated models that allow, for example, for stochastic volatility dynamics and jumps (see, e.g., \citealp{heston1993closed, bates1996jumps, barndorff1997processes, madan1998variance, boyarchenko1999generalizations, boyarchenko2000option}).
These constructions are able to reproduce the observed skew and curvature of the implied volatility smiles in most of the market conditions.
However, calibrating this kind of models is a critical procedure that does not necessarily meet the accuracy and speed required for a successful risk management.

Calibrating an option pricing model consists in iteratively adjusting the model parameters so that the differences between the prices of liquidly-traded options and the corresponding model prices are minimized.
For most pricing models, this optimization problem has a nonconvex objective function (see, e.g., \citealp{mrazek2017calibration,escobar2016parameters} for the Heston model case), so that the feasible region of solutions displays multiple local minima.
As such, selecting a proper optimizer is not trivial and it is a major area of study in finance. 

In order to perform an accurate calibration, it is possible to adopt a number of global optimizers.
Popular examples are given by stochastic search algorithms such as simulated annealing (see, e.g., \citealp{mrazek2014optimization, ondieki2022swaption}), particle swarm optimization (see, e.g., \citealp{yang2012multi}),  differential evolution (see, e.g., \citealp{amici2025multivariateLevy}), or other evolutionary algorithms such as in \citet{hamida2005recovering}.
These are flexible methods that do not need information about the gradient of the objective function, and the convergence of their solutions to the global optimum is nearly independent of the initial values.
However, stochastic search techniques are computationally burdensome due to the large number of searches and iterations required.
This is especially true when dealing with multidimensional asset price models that involve the calibration of several parameters.

Another strand of literature focuses on multistart optimization methods.
These algorithms run local optimizers from a selected set of initial values and choose the best solutions among these local runs
(see applications, e.g., in \citealp{cont2004nonparametric, amici2025multivariate} for jump processes and \citealp{alfeus2020regularization} for the Heston model). 
Most local optimizers are efficient gradient-based algorithms such as gradient descent and conjugate gradient methods (see, e.g., \citealp{dai2016calibration}).
As such, multistart optimization has the core advantage to be fast with respect to stochastic search methods, provided that the number of local runs is not excessively large.
However, whether or not the gradient can be computed depends on the specific problem.
In case the objective function is not differentiable,  
gradient-based algorithms recur to finite difference approximations of the gradient that can be costly and do not guarantee high accuracy.

The calibration of the Heston model is an example in which gradient-based optimization can be problematic.
Both the Heston pricing function and the gradient are recovered via inverse Fourier transform methods that involve numerical integration, which is a source of inaccuracy for the computation of the gradient.
In particular, the integrand in these functions is discontinuous or highly oscillatory for certain combinations of the model parameters (see \citealp{rouah2013heston}).
Thus, calibrating the Heston model and its extensions properly is an open problem that has interested researchers since the introduction of the model until the recent years (see, e.g., \citealp{engelmann2021calibration,romer2020does,chang2021option}, and the aforementioned works). 

    %


Some authors have efficiently dealt with the calibration issues presented by the Heston model.
A noticeable example is represented by the work of \citet{cui2017full}, in which they demonstrate with numerical tests that the Heston model calibration does not get stuck in local minima, therefore avoiding the need to use global optimization techniques.
In addition, the authors recover the analytical gradient of the Heston pricing function, speeding up the computational time.
However, these valuable results can hardly be generalized to multidimensional models or other sophisticated constructions that allow, for example, for jumps.
In addition, the numerical integration required by the Heston pricing function and its gradient can still be computationally expensive for calibrations on large datasets.

In order to deal with the numerical issues encountered in the objective function and increase the computational speed, it is possible to deploy machine learning tools.
In particular, the general approximation theorem for deep neural networks \citep{funahashi1989approximate} provides a theoretical basis for approximating arbitrary functions with relatively simple mathematical operations. 
The so-approximated functions are computationally fast and are not subject to possible discontinuity issues of the original functions. 
These features make neural networks powerful candidates to approximate the Heston-like pricing functions for calibration purposes.
A noticeable example of this kind is that of \citet{liu2019neural}, which also include jumps in the stochastic volatility process.
\citet{dimitroff2018volatility} use a supervised deep convolutional neural network to fit the Heston model to the implied volatility surface.
\cite{bloch2021deep} use deep learning to dynamically evolve the parameters of a stochastic volatility model with an explicit expression to recover the implied volatility smile.

Most notably, speed is the core feature of neural networks.
In addition, as deep learning technologies continue to evolve, GPU hardware is also advancing in terms of architecture and performance.
This mutually reinforcing relationship enables the scale and sophistication of deep learning models to grow, allowing them to handle larger data sizes and complex problems.
The parallel computing power of GPUs provides significant acceleration support for deep learning tasks, making the training and inference process more efficient. 
While GPUs can be used in conjunction with a variety of optimzation techniques (see, e.g., \citealp{ferreiro2020new, han2021gpu, belletti2020tensor}), in the last years they have significantly promoted the application of deep learning in a number of areas including finance.


Because of these reasons, the recent literature on financial model calibration has demonstrated the superior performance of deep learning techniques with respect to traditional optimization methods (see a selected list of works in \citealp{ruf2020neural}).
However, most deep learning methods only concern the creation of a map between model parameters and the output price.
This can be insufficient to obtain an accurate approximation of sophisticated functions.

To tackle this issue, in the spirit of \citet{huge2020differential} we propose a deep differential network (DDN) for the calibration of the Heston model.
Our DDN adds a differentiation layer to the typical structure of a deep neural network.
This layer is given by the first-order partial derivatives of the network output with respect to some of the input parameters, namely the parameters of the stochastic variance process.
In addition, we define the objective function used for the network training as the sum the of squared differences between the output prices of the network and the reference prices plus the sum of the squared differences between the first-order partial derivatives of the output value and the reference partial derivatives.
This procedure produces a more accurate approximation of the pricing function and preserves the computational speed of the typical feedforward neural networks.

Partial derivatives in the objective function of the training are also used in physics-informed neural networks (see e.g., \citealp{raissi2019physics, lo2023training}), and they have been recently applied in financial calibration contexts by, for example, \citet{hoshisashi2023no}.
In this last work the authors set an objective function that embeds derivative constraints in order to ensure that the no-arbitrage condition of the option pricing model is satisfied.
While this imposes that the partial derivatives with respect to the strike and the time to maturity lie in no-arbitrage intervals, in our approach we force all the partial derivatives with respect to the Heston model parameters to roughly match their true values.


The use of partial derivatives represents a significant distinction of our method with respect to the recent papers in neural networks in option pricing reported in \citet{ruf2020neural};
this is the case, for example, of the noticeable work of \citet{liu2019neural}.
In addition, they use a deep neural network to learn implied volatilities, which can be computationally expensive due to the numerical inversion of the Black-Scholes formula.
In contrast, we employ our network to directly learn option prices, applying suitable scaling methods to deal with different orders of magnitudes in prices.


Once the network is trained, we calibrate the Heston model by minimizing the squared differences between market option quotes and the corresponding DDN prices.
As neural networks are ideal constructions for parallel computing algorithms, we can quickly back-calculate the optimal Heston parameters with deep learning-based optimizers.
Therefore, the speed of the DDN lies both in the option valuation and in the calibration procedure, making it extremely faster than most traditional calibration methods.

In addition to the above, we propose a generation of the DDN input dataset via Latin hypercube sampling \citep{mckay2000comparison}, which helps us to better cover the ranges of the parameter values with respect to pseudo-random sampling.
In this way we train the DDN on a huge variety of market conditions, avoiding the need for frequent retraining.

In order to show the power of our DDN, we perform an extensive calibration test on multiple equity markets: the Microsoft, Costco and Booking stocks, and the S\&P500 index.
We compare the performance of the DDN with the performances of the Nelder-Mead method \citep{nelder1965simplex} and of a feedforward neural network that does not embed a differentiation layer, henceforth FNN.
To provide a clean comparison between the calibration techniques, we do not consider options with relatively short maturities (i.e., below one month) and that are relatively far from the at-the-money position (i.e., more than 20\% away from the spot), in which the Fourier-based pricing method may produce nonnegligible errors and affect the quality of the analysis.

The analyzed FNN calibration is conceptually similar to that of \citet{liu2019neural}, albeit with some differences. 
\citet{liu2019neural} use implied volatilities as network outputs and adopt a differential evolution algorithm to mitigate the risk of local minima.
The networks implemented in this paper (i.e., the DDN and the FNN) learn option prices, thereby avoiding the need for numerical inversion of the Black–Scholes formula; to ensure numerical stability, we apply appropriate scaling techniques to normalize price magnitudes.
Furthermore, we employ a multistart optimization strategy to enhance global convergence: the neural network-based optimizer is initialized from multiple randomly generated starting points, and the best-performing solution is retained. This strategy is particularly well-suited to the one-dimensional Heston model, where reasonable parameter bounds may yield an objective function that is approximately convex (see, e.g., \citealp{cui2017full}).
We emphasize that both the choice of global optimization algorithm and the network output (price or implied volatility) are modular components in our framework and do not influence the core comparison between the DDN and the FNN. In this study, the comparison is centered on the presence or absence of the differential layer.

Our results show that the DDN produces a significantly more accurate calibration than the FNN.
Moreover, it increases the calibration speed dramatically with respect to the Nelder-Mead method, preserving roughly the same accuracy.

The remainder of the paper is organized as follows.
In Section \ref{sec:theory} we recall the main theoretical aspects of the Heston model and the possible implementations of the related formulas.
In Section \ref{sec:DDN} we describe the construction of the deep differential network and show how it can be trained.
In Section \ref{sec:empirical_analysis} we report the empirical setting and results, and Section \ref{sec:conclusions} concludes.



\section{The Heston model and its implementations}\label{sec:theory}

In the Heston model \citep{heston1993closed} both the underlying asset price and its variance evolve stochastically over time.
In particular, let $S_t$, $t\geq0$, and $v_t$, $t\geq0$, be the asset price process and the variance process, respectively.
Then, the two risk-neutral dynamics read
\begin{subequations}
\begin{eqnarray}
    &&\dd S_t = r S_t \:\dd t + \sqrt{v_t} S_t \:\dd W_{1,t}
    \\
    &&\dd v_t = \kappa (\lambda - v_t) \:\dd t + \sigma \sqrt{v_t} \:\dd W_{2,t}
    \\
    &&\mathbb{E}^\mathbb{Q} [\dd W_{1,t} \: \dd W_{2,t}] = \rho \:\dd t,
\end{eqnarray}
\end{subequations}
where $W_{1,t}$, $t\geq0$, and $W_{2,t}$, $t\geq0$, are mutually correlated $\mathbb{R}$-valued Brownian motions, $r$ is the continuously compounded risk-free interest rate (assumed to be constant), $\mathbb{Q}$ is an equivalent martingale measure, and 
\begin{equation}\label{eq:Heston_parameters}
\boldsymbol{\theta}_\mathrm{H} = \left(
\kappa,\lambda,\sigma,\rho,v_0
\right)
\end{equation}
is the vector of model parameters, not observable in the market.
In particular, $\kappa>0$ is the mean reversion speed of the variance process, $\lambda>0$ is the long run mean of the variance, $\sigma>0$ is the volatility of the variance, $\rho\in[-1,1]$ drives the correlation between the stock price and the variance, and $v_0>0$ is the initial value of the variance.

The author provides a semi-analytical formulation for pricing plain-vanilla options based on the inverse Fourier transform.
Let $K$ and $\tau$ be the strike price and the time to maturity of a call option, respectively, and let the parameter vector $\boldsymbol{\theta}$ be defined as
\begin{equation}\label{eq:theta}
\boldsymbol{\theta} =
\left(
\kappa,\lambda,\sigma,\rho,v_0,
S_0, r, \tau, K
\right).
\end{equation}
Then, the time-0 valuation of a call option under the Heston model reads
\begin{equation}\label{eq:heston_oushi}
\begin{aligned}
\mathcal{G} \left( \tt \right)
& = \ee^{-r\tau}
\mathbb{E}^{\mathbb{Q}}
\left[(S_\tau - K)^+\right] =S_0 \Pi_1-K \ee^{-r \tau} \Pi_2
\end{aligned}
\end{equation}
where
\begin{subequations}
\begin{eqnarray}
   &&\Pi_1 = \frac{1}{2}+\frac{1}{\pi} \int_0^{\infty} \operatorname{Re}\left(\frac{\phi_\tau(u-\mathrm{i})}{\mathrm{i} u} \mathrm{e}^{-\mathrm{i} k u}\right) \mathrm{d} u,
\\
    &&\Pi_2 =\frac{1}{2}+\frac{1}{\pi} \int_0^{\infty} \operatorname{Re}\left(\frac{\phi_\tau(u)}{\mathrm{i} u} \mathrm{e}^{-\mathrm{i} k u}\right) \mathrm{d} u,
\end{eqnarray}
\end{subequations}
$\phi_\tau(u)$, $u\in\mathbb{R},$ denotes the characteristic function of 
$x_{\tau}\coloneqq \log(S_{\tau}/S_0)$, Re($\cdot$) returns the real part of a complex number, and $k \coloneqq \log(K/S_0)$.
Although our focus is on call options, using Eq.\ \eqref{eq:heston_oushi} it is easy to recover the plain-vanilla put option price via put-call parity as $\mathcal{P} \left( \tt \right) = \GG \left( \tt \right)- S_0 + K \mathrm{e}^{-r \tau}$.

In order to overcome the branch-cut issues of the Heston characteristic function (see a discussion in \citealp{albrecher2007little}) it is possible to express the characteristic function (provided, for example, in \citealp{schoutens2003perfect} and \citealp{gatheral2006volatility}) that reads
\begin{equation}\label{eq:affine_CF}
    \phi_\tau(u)=\exp \left(C_\tau(u)+D_\tau(u) v_0+\mathrm{i} u x_0\right)
\end{equation}
where
\begin{equation}\label{eq:CD}
\begin{aligned}    
&C_\tau(u) =\ii r u \tau+\frac{\kappa \lambda}{\sigma^2}
\left(
\left(
\kappa-\ii \rho\sigma u -d(u)\right) \tau- 
2 \log \left(\frac{1-g(u) \ee^{-d(u) \tau}}{1-g(u)}
\right)
\right),
\\
&D_\tau(u) = \frac{\kappa-\ii \rho \sigma u - d(u)}{\sigma^2}\left(\frac{1-\ee^{-d(u) \tau}}{1-g(u) \ee^{-d(u) \tau}}\right),
\\
&g(u) = \frac{\kappa-\ii \rho \sigma u - d(u)}{\kappa-\ii \rho \sigma u + d(u)},
\hspace{0.4cm}
d(u) = \sqrt{(\kappa-\ii \rho \sigma u)^2+\sigma^2\left(\ii u+u^2\right)}.
\end{aligned}
\end{equation}

A number of Fourier-based techniques are available to recover the Heston price of the call option (see, e.g., \citealp{carr1999option, lewis2000option, fang2009novel}, and \citealp{QuantLibReference} for an efficient code implementation).
Also, \citet{cui2017full} derive the expression for the semi-analytical gradient, for the computation of which it is still possible to apply the aforementioned Fourier techniques.
In this work, we generate the dataset used for the neural network training by pricing the call options through the QuantLib engine,
which efficiently approximates the integrals of the pricing function by means of exponentially-fitted Gauss-Laguerre quadrature in the Lewis framework (see \citealp{QuantLibReference-Heston} for details of the implementation).\footnote{In Appendix \ref{app:QuantLib}, the interested reader can observe the Heston call option prices computed with the \citet{QuantLibReference} pricer and the corresponding partial derivatives for selected combinations of parameters obtained from the calibrations of Section \ref{sec:calibration_results}.}
In order to be consistent with the QuantLib pricing methodology, we recover the partial derivatives by numerical differentiation of the QuantLib pricer, given that the semi-analytical partial derivatives are not available in QuantLib.
In addition, working with the numerical partial derivatives makes our analysis more generalizable (e.g., in terms of computational time) to other complex pricing frameworks, in which analytical differentiation is often not possible.

\section{Deep differential network}\label{sec:DDN}
In this section we introduce our deep differential network (DDN) and describe how it can be trained and used for calibration purposes.

\subsection{Structure}\label{sec:structure}
Our network preserves most of the characteristics of a typical feedforward neural network.
Thus, it consists on a number of layers, each including a set of nodes.
Let $L$ be the number of layers and $N_l$ be the number of nodes of the $l$-th layer.
Then, the values of the nodes of the $l$-th layer are initially computed as
\begin{equation}\label{eq:x^l}
    \xx^{(l)} = \WW^{(l)} \yy^{(l-1)} + \bb^{(l)},
\end{equation}
where $\WW^{(l)} \in \mathbb{R}^{N_l \times N_{l-1}}$ is a matrix of weights, $\bb^{(l)} \in \mathbb{R}^{N_l}$ is a bias vector, and $\yy^{(l-1)}$ is the value of the $(l-1)$-th node vector.
In order to introduce nonlinearity in the network, the $l$-th layer is subsequently modified by means of a nonlinear function $\psi_l: \mathbb{R}^l \rightarrow \mathbb{R}^l$, so that
\begin{equation}\label{eq:y^l}
    \yy^{(l)} = \psi_l \left(\xx^{(l)}\right)
\end{equation}
is the \lq\lq activated\rq\rq\ value of the $l$-th node vector.
Eqs.\ \eqref{eq:x^l} and \eqref{eq:y^l} describe how the input vector is forward propagated to the output.

\begin{figure} 
    \centering
    \includegraphics[width=0.65\textwidth]{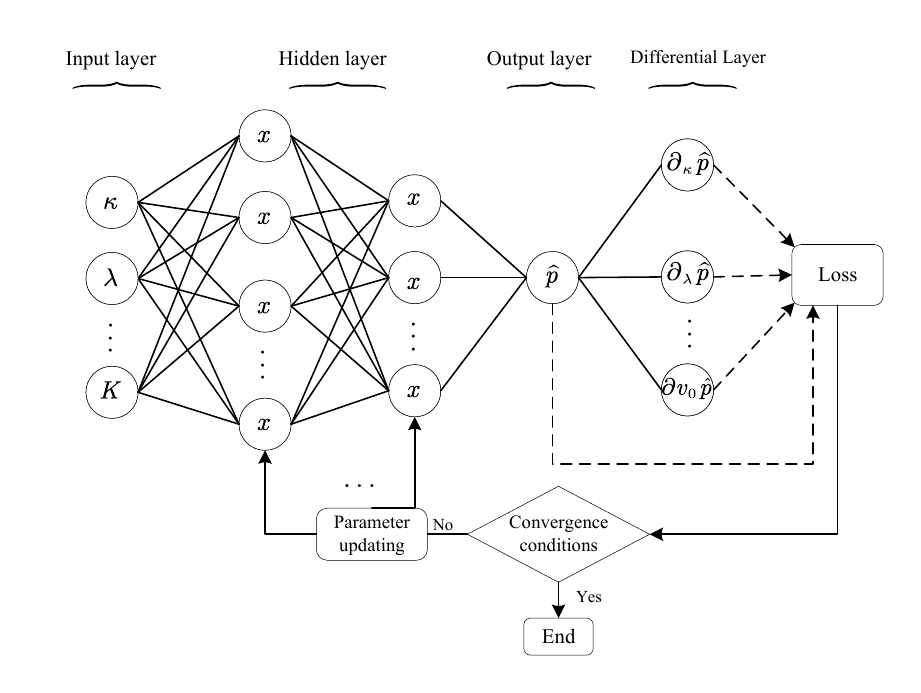}
    \caption{
    Topology of the deep differential network and training process scheme.
    }
    \label{fig:DDN}
\end{figure}

While the number and the dimension of the middle, or hidden, layers is arbitrary, the input and the output layers are defined by the specific problem.
In our network the input layer is represented by the parameter vector $\boldsymbol{\theta} \in \mathbb{R}^\mathcal{I}$, which includes the Heston model parameters $\kappa, \lambda, \sigma, \rho, v_0$, and the observable data $S_0$, $r$, $\tau$, and $K$, 
so that $\mathcal{I}=9$.
The output layer only contains the option price and is calculated with the network predictor $f(\tt)$.
In addition, we design a differentiation layer in which we compute the first-order partial derivatives of the output with respect to the five input nodes that represent the Heston parameters.
The diagram of our deep differential network is presented in Figure \ref{fig:DDN}.
While the calculation of the output layer $f(\tt) = \yy^{(L)}$ follows from Eqs.\ \eqref{eq:x^l} and \eqref{eq:y^l}, the differentiation layer $\frac{\partial f(\boldsymbol{\theta})}{\partial \boldsymbol{\theta}_{\mathrm{H}}}$ is recovered by selecting the first five entries of the gradient
\begin{equation}\label{eq:df_dtheta}
\begin{aligned}
\frac{\partial f(\boldsymbol{\theta})}{\partial \boldsymbol{\theta}}
& = \frac{\partial f(\boldsymbol{\theta})}{\partial \yy^{(L)}} \frac{\partial \yy^{(L)}}{\partial \xx^{(L)}} \frac{\partial \xx^{(L)}}{\partial \yy^{(L-1)}} \cdots \frac{\partial \xx^{(1)}}{\partial \yy^{(0)}}
\\
& =
\psi_L^{\prime}\left(\xx^{(L)}
\right) \WW^{(L)} \cdots \operatorname{diag}\left(\psi_1^{\prime}\left(\xx^{(1)}\right)\right) \WW^{(1)},
\end{aligned}
\end{equation}
where $\psi^{\prime}(\cdot)$ denotes the derivative of $\psi(\cdot)$.

\subsection{Training process}\label{sec:training_theory}
In order to train the network to a specific dataset, we need to search for the weights and biases that optimize a selected objective function.

\paragraph{Objective function}
In our objective function, we minimize the differences between the predicted prices and the Heston prices, and between the predicted partial derivatives and the numerical partial derivatives computed over the Heston prices.
Let $p = \GG(\tt)$ denote the call option price computed with the Heston formula and $\hat{p} = f(\tt)$ be the corresponding prediction of the network pricer $f$.
Also, let $\boldsymbol{\Xi}$ be the stacked vector of all the weights and biases of the network, and $B < N$ be a selected batch size of the training data, where $N$ is the total number of training samples.
Then, we set the total objective function $\mathcal{J}$ as
\begin{subequations}
\begin{eqnarray}
    && \mathcal{J}(\mathcal{L}; \boldsymbol{\Xi})
    = \mathcal{L}_1 
    \left( \hat{\pp}, \pp \right) +
    \mathcal{L}_2 \left( \boldsymbol{\dd}_{\boldsymbol{\theta}_\mathrm{H}} \hat{\pp}, \boldsymbol{\dd}_{\boldsymbol{\theta}_\mathrm{H}} \pp \right),
    \\
    &&
    \hat{\pp} = \left( \hat{p}^{(1)},
    \cdots, \hat{p}^{(n)},
    \cdots, \hat{p}^{(B)} \right),
    \quad
    \pp = \left( p^{(1)},
    \cdots, p^{(n)},
    \cdots, p^{(B)} \right),
    \\
    &&
    \boldsymbol{\dd}_{\tt_\mathrm{H}} \hat{\pp} = \left( \partial_{\tt_\mathrm{H}} \hat{p}^{(1)},
    \cdots, \partial_{\tt_\mathrm{H}} \hat{p}^{(n)},
    \cdots, \partial_{\tt_\mathrm{H}} \hat{p}^{(B)} \right)\!,
    \\
    &&\boldsymbol{\dd}_{\tt_\mathrm{H}} \pp = \left( \partial_{\tt_\mathrm{H}} p^{(1)},
    \cdots, \partial_{\tt_\mathrm{H}} p^{(n)},
    \cdots, \partial_{\tt_\mathrm{H}} p^{(B)} \right)\!,
\end{eqnarray}
\end{subequations}
where
 $\mathcal{L} = \{\mathcal{L}_1, \mathcal{L}_2\}$ is defined according to the chosen loss measures, and $p^{(n)}$ and $\hat{p}^{(n)}$ are the Heston and the DDN prices of the $n$-th training sample, respectively.

A common practice to avoid network overfitting is to regularize the objective function.
Thus, we let the objective function include a penalty term and be redefined as
\begin{equation}\label{eq:cost_function_reg}
 \mathcal{R}(\mathcal{L}, \boldsymbol{\Xi})
 = \mathcal{J}(\mathcal{L}, \boldsymbol{\Xi}) + \eta \norm{\boldsymbol{\Xi}}_{2}, 
\end{equation}
where $\eta$ is a regularization coefficient.



\paragraph{Update of the network parameters}
The optimization of the objective function of a neural network is typically performed via gradient-based methods.
In particular, the weights and biases of Eq.\ \eqref{eq:x^l} are updated as
\begin{equation}\label{eq:canshugengxin_W}
\begin{aligned}
\boldsymbol{W}^{(l)} \gets \boldsymbol{W}^{(l)}-\alpha \left( \frac{\partial \mathcal{R}(\mathcal{L}, \boldsymbol{\Xi})}{\partial \boldsymbol{W}^{(l)}} \right)^{\top}, 
\end{aligned}
\end{equation}
\begin{equation}\label{eq:canshugengxin_b}
    \boldsymbol{b}^{(l)} \gets \boldsymbol{b}^{(l)}-\alpha \left( \frac{\partial \mathcal{R}(\mathcal{L}, \boldsymbol{\Xi})}{\partial \boldsymbol{b}^{(l)}} \right)^{\top},
\end{equation}
where $\alpha$ is a selected hyperparameter known as the learning rate.

In order to calculate the partial derivatives in Eqs.\ \eqref{eq:canshugengxin_W} and \eqref{eq:canshugengxin_b} we first note that
%
\begin{eqnarray}
    \frac{\partial \mathcal{R}(\mathcal{L}, \boldsymbol{\Xi})}{\partial w_{ji}^{(l)}}
    &=&
    \frac{\partial \mathcal{R}(\mathcal{L}, \boldsymbol{\Xi})}{\partial \boldsymbol{x}^{(l)}}
    \frac{\partial \boldsymbol{x}^{(l)}}{\partial w_{ji}^{(l)}},
    \\
    \frac{\partial \mathcal{R}(\mathcal{L}, \boldsymbol{\Xi}) }{\partial b_j^{(l)}}
    &=&
    \frac{\mathcal{R}(\mathcal{L}, \boldsymbol{\Xi})}{\partial \boldsymbol{x}^{(l)}}
    \frac{\partial \boldsymbol{x}^{(l)}}{\partial b_j^{(l)}}.
\end{eqnarray}
%
The terms on the right-hand sides of the above equations can be rearranged and computed as follows.
First, we have
\begin{equation}\label{delta_p}
\begin{aligned}
\frac{\partial \mathcal{R}(\mathcal{L}, \boldsymbol{\Xi})}{\partial \boldsymbol{x}^{(l)}}
& =  \frac{\partial \mathcal{R}(\mathcal{L}, \boldsymbol{\Xi})}{\partial \boldsymbol{y}^{(l)}}  \frac{\partial \boldsymbol{y}^{(l)}}{\partial \boldsymbol{x}^{(l)}}  \\
& =  \frac{\partial \mathcal{R}(\mathcal{L}, \boldsymbol{\Xi})}{\partial \boldsymbol{x}^{(l+1)}} \frac{\partial \boldsymbol{x}^{(l+1)}}{\partial \boldsymbol{y}^{(l)}}  \frac{\partial \boldsymbol{y}^{(l)}}{\partial \boldsymbol{x}^{(l)}}  \\
& =\frac{\partial \mathcal{R}(\mathcal{L}, \boldsymbol{\Xi})}{\partial \boldsymbol{x}^{(l+1)}} \boldsymbol{W}^{(l+1)} \operatorname{diag}\left(\psi_{l}^{\prime}\left(\boldsymbol{x}^{(l)}\right)\right) \\
& =
\left( \frac{\partial \mathcal{R}(\mathcal{L}, \boldsymbol{\Xi})}{\partial \boldsymbol{x}^{(l+1)}}   \boldsymbol{W}^{(l+1)}\right) \odot\psi_{l}^{\prime}\left(\boldsymbol{x}^{(l)}\right)
:= \boldsymbol{\zeta}^{(l)},
\\
\end{aligned}
\end{equation}
where $\odot$ denotes element-wise multiplication.
For $l=L$, the value of $\frac{\partial \mathcal{R}(\mathcal{L}, \boldsymbol{\Xi})}{\partial \yy^{(l)}} = \frac{\partial \mathcal{R}(\mathcal{L}, \boldsymbol{\Xi})}{\partial \hat{\pp}}$ is known and can be used to recursively get all the solutions of the form \eqref{delta_p}.
Moreover,

\begin{equation} 
    \begin{aligned}
    \frac{\partial \boldsymbol{x}^{(l)}}{\partial w_{ji}^{(l)}}
    & =\left(\frac{
    \partial x_1^{(l)}}{\partial w_{ji}^{(l)}}, \cdots, \frac{\partial x_j^{(l)}}{\partial w_{ji}^{(l)}}, \cdots, \frac{\partial x_{N_l}^{(l)}}{\partial w_{ji}^{(l)}}\right)^{\top} \\
    & =\left(0, \cdots, \frac{\partial\left(\boldsymbol{w}_{j:}^{(l)} \boldsymbol{y}^{(l-1)}+b_j^{(l)}\right)}{\partial w_{ji}^{(l)}}, \cdots, 0\right)^{\top} \\
    & =\left(0, \cdots, \underbrace{y_i^{(l-1)}}_{j^\text{th}}, \cdots, 0\right)^{\top}, \\
    \frac{\partial \boldsymbol{x}^{(l)}}{\partial b_{j}^{(l)}}
    & =\left(0, \cdots, \underbrace{1}_{j^\text{th}}, \cdots, 0\right)^{\top}, \\
\end{aligned}
\end{equation}
where $\boldsymbol{w}_{j:}^{(l)}$ is the $j$-th row of the weight matrix $\boldsymbol{W}^{(l)}$.
%
%
As a result, it is easy to show that the partial derivatives in Eqs.\ \eqref{eq:canshugengxin_W} and \eqref{eq:canshugengxin_b} are given by
\begin{equation}
    \frac{\partial \mathcal{R}(\mathcal{L}, \boldsymbol{\Xi})}{\partial \boldsymbol{W}^{(l)}}
 = \boldsymbol{y}^{(l-1)}  \boldsymbol{\zeta}^{(l)},\qquad\text{and}\qquad
    \quad \frac{\partial \mathcal{R}(\mathcal{L}, \boldsymbol{\Xi}) }{\partial \boldsymbol{b}^{(l)}} = \boldsymbol{\zeta}^{(l)},
\end{equation}
respectively.



For each batch of the training dataset,
we compute the DDN output and the gradient of the network output, recalculate the objective function
and update $\boldsymbol{\Xi}$ according to the backpropagation scheme of Eqs.\ \eqref{eq:canshugengxin_W} and \eqref{eq:canshugengxin_b}.
The procedure is repeated multiple times (as many epochs are set in the training process) until the objective function is minimized and the DDN is sufficiently trained.

\subsection{Calibration scheme}
Once the network $f(\boldsymbol{\theta} \mid \boldsymbol{\Xi})$ is trained, we can use it to approximate the Heston pricing function $\GG(\boldsymbol{\theta})$.
This DDN pricer can be used for a number of purposes, including calibration to market quotes.
Let $p^{\text{mkt}}_1, \ldots, p^{\text{mkt}}_M$ be the market prices of $M$ exchange-traded options.
Then, the calibration problem can be designed as
\begin{equation}\label{get_parameters}
\begin{aligned}
    \boldsymbol{\theta}_\mathrm{H}^* =
    \hspace{7pt}
    & \underset{\boldsymbol{\theta}_\mathrm{H} \in \overline{\boldsymbol{\theta}}_\mathrm{H}}{\operatorname{argmin}} && \frac{1}{M} \sum_{m=1}^{M} \left( f_m(\boldsymbol{\theta} \mid \boldsymbol{\Xi}) - p^{\text{mkt}}_m \right)^2,
\end{aligned}
\end{equation}
where $\overline{\boldsymbol{\theta}}_\mathrm{H}$ is a feasible region of solutions for the Heston parameters $\boldsymbol{\theta}_\mathrm{H}$, and $f_m$ denotes the network pricer with strike and maturity given by the $m$-th option.
We remark once again that the DDN pricer $f(\boldsymbol{\theta} \mid \boldsymbol{\Xi})$ allows for an easy extraction of the gradient, due to its neural network-based structure.
As such, we can solve the calibration problem with fast gradient-based algorithms that would otherwise risk to produce numerical issues if the pricing function in Eq.\ \eqref{get_parameters} was the Heston formula of Eq.\ \eqref{eq:heston_oushi}.

\section{Empirical analysis}\label{sec:empirical_analysis}


In this section we describe our empirical tests and show the results that demonstrate the validity of our calibration method based on the deep differential network.

\subsection{Data generation and preprocessing}\label{sec:preprocessing}

In order to generate the inputs of the dataset, we use the Latin hypercube sampling (LHS) technique (see \citealp{mckay2000comparison}). 
As the authors show, the LHS ensures an efficient generation of the space of interest by dividing it into equivalent intervals and sampling from each interval exactly once.
This allows one to cover the input space with fewer samples than what pseudo-random numbers would require, avoiding to undersample some regions.

We do not impose the Feller condition in the generation of parameter instances. As noted, for example, by \citet{pacati2018smiling}, enforcing this constraint can hinder the ability of the model to fit market data accurately. Moreover, we verify the robustness of our methodology both with and without applying this constraint, and consequently choose to relax the Feller condition in order to enhance model flexibility.

It is worth mentioning that whatever data generation method is used, some combinations of the input parameters could hardly correspond to realistic volatility surfaces; however, in order to make the DDN thoroughly learn the Heston pricing function, we keep a sufficiently wide variety of input parameters.

The ranges of the parameters provided to the LHS engine are reported in Table \ref{tab:Generating}, where the strike price $K$ is first generated in terms of $\log \left(K / S_0\right)$ and then suitably rescaled.
Then, the corresponding dataset outputs are directly obtained by applying the Heston pricing formula \eqref{eq:heston_oushi} to each parameter combination generated via LHS.
We compute the Heston prices and the corresponding numerical gradients via the \citet{QuantLibReference} package.



\begin{table} 
\centering
\caption{
Ranges of the input parameters of the network dataset.
Heston model parameters in Eq.\ \eqref{eq:Heston_parameters}.
$S_0$ = initial price of the underlying asset.
$r$ = risk-free interest rate.
$\tau$ = time to maturity of the option.
$K$ = strike price.
}
\begin{tabular}{cl}
\hline
Parameter & Range  \\
\hline
$\kappa$ & {$[0.005,5]$}  \\
$\lambda$ & {$[0,1]$}  \\
$\sigma$ & {$[0.1,1]$}  \\
$\rho$ & {$[-0.95,0]$}  \\
$v_0$ & {$[0,1]$}  \\
$r$ & {$[0,0.10]$}  \\
$\tau$ & {$[0.05,1]$}  \\
$S_0$ & {$[10,6000]$}  \\
$\log\left(K / S_0\right)$ & {$[-1,1]$}  \\
\hline
\end{tabular}
\label{tab:Generating}
\end{table}


Furthermore, in order to eliminate the scaling differences between data and reduce the influence of outliers, 
we normalize the features and the labels of the network, along with the values of the differential layer.
This guarantees a stable training of the DDN.
Thus, we modify the input data as
\begin{equation}\label{eq:normal_X}
\begin{aligned}    
    & \Tilde{\theta}_{i}^{(n)}
    = \frac{
    \theta_{i}^{(n)} - \min_n \left(\theta_{i}^{(n)} \right)
    }{
    \max_n \left(\theta_{i}^{(n)} \right) - \min_n \left(\theta_{i}^{(n)} \right)
    },
    \quad i=1, \ldots, \mathcal{I},
    \quad n=1, \ldots, N^*,
    \\
    & \Tilde{\hat{p}}^{(n)}
    = \frac{
    \hat{p}^{(n)} - \min_n \left(\hat{p}^{(n)} \right)
    }{
    \max_n \left(\hat{p}^{(n)} \right) - \min_n \left(\hat{p}^{(n)} \right)
    },
    \quad n=1, \ldots, N^*,
\end{aligned}
\end{equation}
where $\mathcal{I}$ is the dimension of the input layer and $N^*$ is the number of data points.
%
We then obtain the standardization of the first-order partial derivatives as follows.
Set $\delta_{\theta_i} = \max_n \left( \theta^{(n)}_{i} \right) - \min_n \left( \theta^{(n)}_{i} \right)$, $\delta_{p} = \max_n \left( \hat{p}^{(n)} \right) - \min_n \left( \hat{p}^{(n)} \right)$, then
\begin{equation}
   \begin{aligned}
    \frac{\partial \Tilde{\hat{p}}^{(n)}}{\partial \Tilde{\theta}_j^{(n)}}
    & =
    \frac{\partial \Tilde{\hat{p}}^{(n)}}{\partial \hat{p}^{(n)}}
    \frac{\partial \hat{p}^{(n)}}{\partial \theta_j^{(n)}}
    \frac{\partial \theta_j^{(n)}}{\partial \Tilde{\theta}_j^{(n)}}
    \\
    & =
    \frac{\delta_{\theta_i}}{\delta_{p}}
    \frac{\partial \hat{p}^{(n)}}{\partial \theta_j^{(n)}},
    \quad i = 1, \ldots, \mathcal{I}_H,
    \quad n = 1, \ldots, N^*,
\end{aligned} 
\label{eq:partial_X/Y}
\end{equation}
where $\frac{\partial \hat{p}^{(n)}}{\partial \theta_j^{(n)}}$ is the corresponding unnormalized partial derivative computed as in Eq.\ \eqref{eq:df_dtheta}, and $\mathcal{I}_H$ is the number of Heston parameters.

\subsection{Hyperparameters}
In order to suitably train the DDN and prepare it for the calibrations of the next sections, we choose a set of network hyperparameters that we report in Table \ref{tab:hyperparameter}.
Our selection of these hyperparameters undergoes a twofold strategy.
We choose most of them consistently with past works that apply neural networks in a similar context (see, e.g., \citealp{ferguson2018deeply}).
In order to address common issues such as gradient explosion and disappearance, we apply the Xavier Glorot initialization (see \citealp{glorot2010understanding}). Additionally, in order to prevent overfitting during the training stage we regularize the network by dropping out nodes from each layer, setting a dropout rate of 0.2.

However, we select the number of hidden layers, the layer dimensions, and the dataset size according to a number of empirical tests, which we describe in this section.
%
%
In all our experiments, we use an Ubuntu Linux operating system and an NVIDIA GeForce RTX 3060 Laptop GPU. Our network implementation is carried out using the PyTorch framework in Python 3.9.16, with the Spyder IDE as the development environment.

\begin{figure} 
    \centering
    \begin{minipage}{0.3\textwidth}
        \centering
        \includegraphics[width=\linewidth]{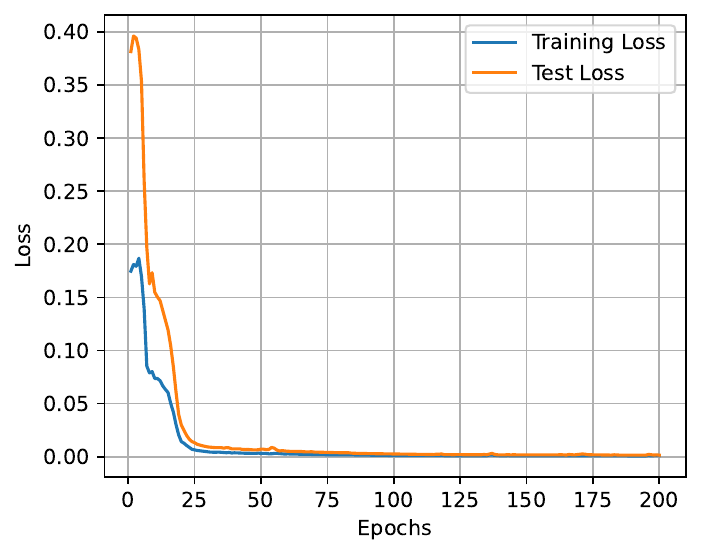}
        \subcaption*{\qquad (a)} 
    \end{minipage}%
    \begin{minipage}{0.3\textwidth}
        \centering
        \includegraphics[width=\linewidth]{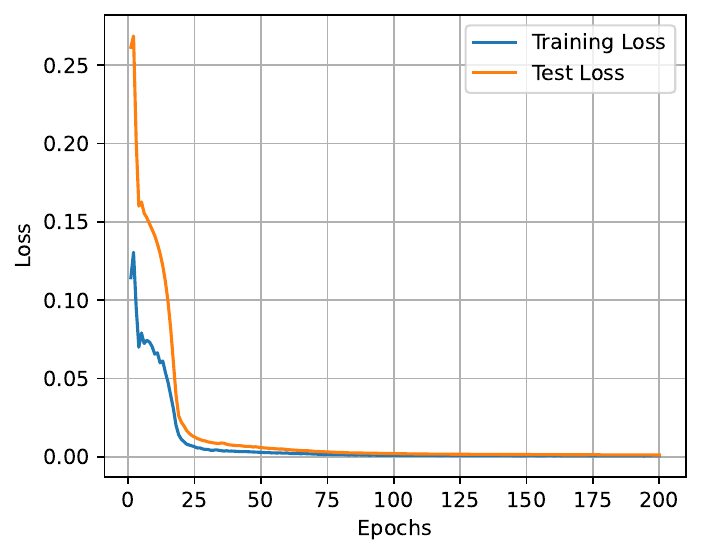}
        \subcaption*{\qquad (b)} 
    \end{minipage}%
    \begin{minipage}{0.3\textwidth}
        \centering
        \includegraphics[width=\linewidth]{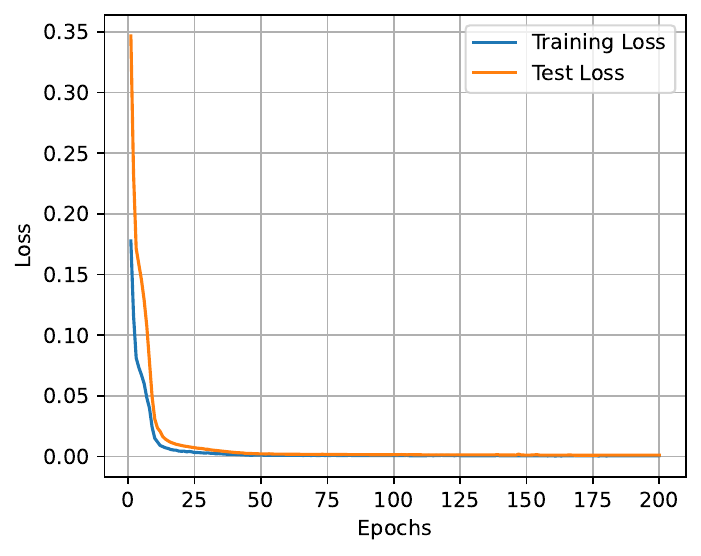}
        \subcaption*{\qquad (c)} 
    \end{minipage}

    \vspace{0.5cm} 

    \begin{minipage}{0.3\textwidth}
        \centering
        \includegraphics[width=\linewidth]{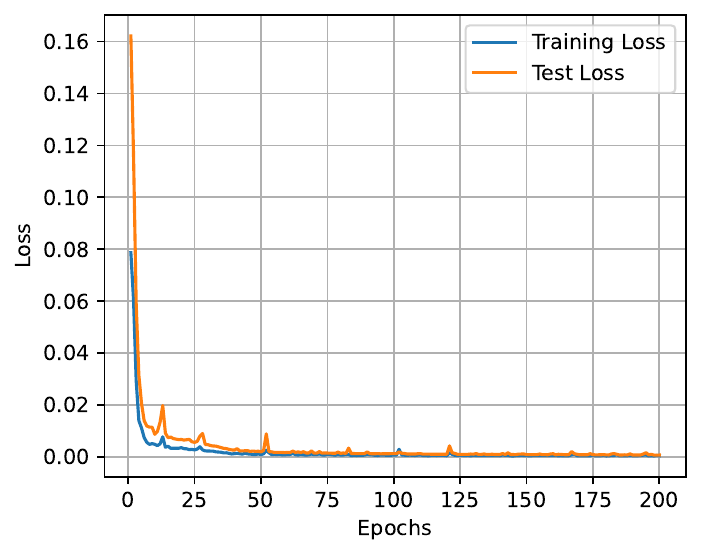}
        \subcaption*{\qquad (d)} 
    \end{minipage}%
    \begin{minipage}{0.3\textwidth}
        \centering
        \includegraphics[width=\linewidth]{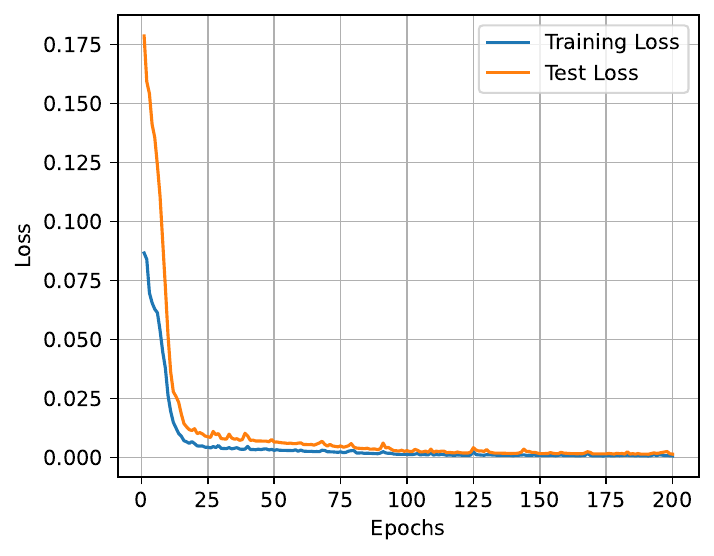}
        \subcaption*{\qquad (e)} 
    \end{minipage}%
    \begin{minipage}{0.3\textwidth}
        \centering
        \includegraphics[width=\linewidth]{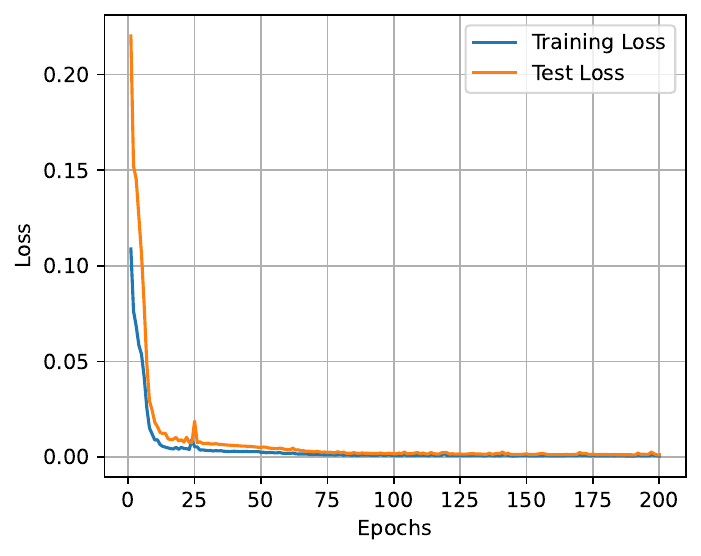}
        \subcaption*{\qquad (f)} 
    \end{minipage}
    \caption{Training and test errors of the six optimal DDN configurations from Table \ref{tab:test data error in Different hidden} (panel (a) corresponds to the 3-layer case, panel (b) to the 4-layer case, and so on), trained on a 10k-sample dataset with a 70:15:15 training, validation, and test split.
}
    \label{fig:DDN TARINING LOSS}
\end{figure}

In order to determine the optimal number of layers and nodes, we initially train the network to a parsimonious dataset of 10k samples under different network topologies and compare the errors of the test sets.
Table \ref{tab:test data error in Different  hidden} shows the optimal number of nodes per layer for several depth levels,
and Figure \ref{fig:DDN TARINING LOSS} displays the training and test set errors of the six optimal cases of Table \ref{tab:test data error in Different  hidden} as functions of the number of epochs.
Interestingly, simply increasing the number of hidden layers or the number of neurons does not necessarily improve the accuracy of the network. 
In terms of overall loss, the case with 6 hidden layers and 150 neurons per layer ouperforms the other configurations, which is why we choose it for our next tests.

%
%


\begin{table} 
    \centering
        \caption{
        Selected hyperparameters of the DDN.
        Adam optimizer introduced by \citet{kingma2015method}.
        Softplus activation function defined as $\psi(x) = \log (1+\exp (\beta x))$.
        MSE: mean square error.
        The decay rate regulates the level of the learning rate over the different epochs.
        }
\begin{tabular}{ll}
\hline Hyperparameter & Choice \\
\hline
 Optimization algorithm & Adam \\
 Initial learning rate & 0.001 \\
 Decay rate & 0.9 \\
 Number of hidden layers & 6 \\
 Number of neurons per hidden layer & 150 \\
 Activation function & Softplus \\
 Number of epochs & 200 \\
 Loss function & MSE \\
 Dataset size & 200k \\
 Training/validation/test set ratio & 70:15:15 \\
 Batch size & 256 \\
 Dropout & 0.2\\
\hline
\end{tabular}
\label{tab:hyperparameter}
\end{table}


\begin{table} 
\centering
\caption{
Test set errors under different DDN configurations.
DDN trained and tested on a dataset of 10k samples with a 70:15:15 ratio between training, validation and test set.
Bold numbers represent the minimum error of each row.
}
\begin{tabular}{ccccc}
\cline{2-5}
\multicolumn{1}{c}{} & \multicolumn{4}{c}{Number of nodes per hidden layer}
\\
\hline Number of hidden layers & 50 & 100 & 150 & 200  \\
\hline
 3 & $4.03 \times 10^{-3}$ & $\boldsymbol{3.64 \times 10^{-3}}$ & $3.36 \times 10^{-3}$ & $3.87 \times 10^{-3}$ \\
 4 & $4.06 \times 10^{-3}$ & $3.84 \times 10^{-3}$  & $3.92 \times 10^{-3}$ & $\boldsymbol{3.62 \times 10^{-3}}$ \\
 5 & $3.77 \times 10^{-3}$ & $3.72 \times 10^{-3}$ & $\boldsymbol{3.42 \times 10^{-3}}$ & $3.61 \times 10^{-3}$ \\
 6 & $3.68 \times 10^{-3}$ & $3.74 \times 10^{-3}$ &$\boldsymbol{3.33 \times 10^{-3}}$ & $3.82 \times 10^{-3}$ \\
  7 & $3.54 \times 10^{-3}$ & $\boldsymbol{3.44 \times 10^{-3}}$ &$3.53 \times 10^{-3}$ & $3.64 \times 10^{-3}$ \\
   8 & $\boldsymbol{3.68 \times 10^{-3}}$ & $3.74 \times 10^{-3}$ &$3.83 \times 10^{-3}$ & $3.81 \times 10^{-3}$ \\
\hline
\end{tabular}
\label{tab:test data error in Different  hidden}
\end{table}

\begin{figure} 
    \centering
    \includegraphics[width=6.5cm]{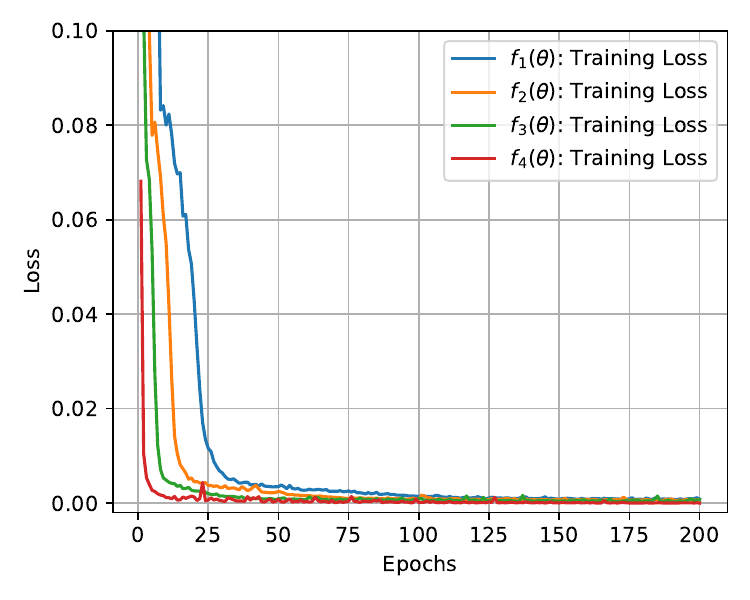}
    \includegraphics[width=6.5cm]{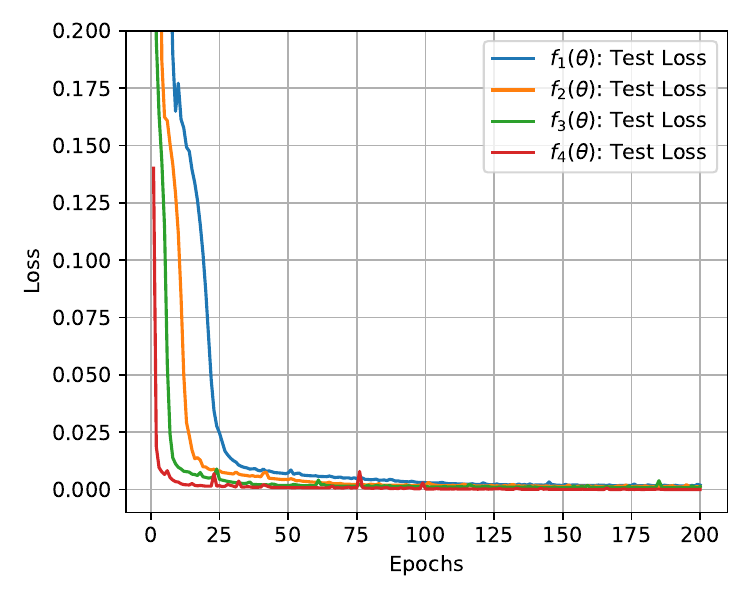}
    \caption{
        Training and test loss of the DDN (as a function of the number of epochs) for 10k, 50k, 100k, and 200k samples.    
        }
    \label{fig:DDN LOSS IN TRAINING AND TESTING}
\end{figure}

In order to select the dataset size, we compare the performance of networks trained and tested on four different datasets.
We construct them by generating 10k, 50k, 100k, and 200k samples, respectively.
%
We train and test the datasets using the hyperparameters of Table \ref{tab:hyperparameter}.
Not surprisingly, larger samples result in smaller training errors, as reported by Table \ref{tab:four approximation} and by Figure \ref{fig:DDN LOSS IN TRAINING AND TESTING}, in which we plot the training and validation loss curves.
In order to use a network that performs well out of sample, we choose the dataset size according to the results on the test set.
As the DDN enjoys the best performance with the dataset of 200k samples, we set this dataset size for the rest of our empirical exercises.

We further point out that, as shown in the last column of Table \ref{tab:four approximation}, the DDN training times roughly correspond to the training times of a typical feedforward neural network in which the differentiation layer is not employed, remarking the parsimony of our methodology.



\begin{table} 
\centering
\caption{
Training, validation, and test set errors of the DDN on datasets of different sizes.
Last column: training times of the DDN; in brackets, training times of a feedforward neural network that does not embed the differentiation layer.
}
\begin{tabular}{cccccc}
\cline{2-6} & Dataset size & Training loss & Validation loss & Test loss & Training time\\
\hline  $f_1(\boldsymbol{\theta})$  & $10$k& $2.37 \times 10^{-3}$ & $3.02 \times 10^{-3}$ & $3.33 \times 10^{-3}$ &84s (76s)\\
 $f_2(\boldsymbol{\theta})$ & $50$k & $9.24 \times 10^{-4}$ & $1.86 \times 10^{-3}$ & $2.94 \times 10^{-3}$ & 10m31s (9m42s)\\
 $f_3(\boldsymbol{\theta})$ & $100$k  & $3.62 \times 10^{-4}$ & $4.03 \times 10^{-4}$ & $4.64 \times 10^{-4}$ &  45m35s (43m24s)\\
 $f_4(\boldsymbol{\theta})$   & $200$k & $1.32 \times 10^{-5}$ & $3.26 \times 10^{-5}$ & $4.24 \times 10^{-5}$ & 2h32m (2h28m)\\
\hline
\end{tabular}
\label{tab:four approximation}
\end{table}




\subsection{Calibration setting}\label{sec:calibration_setting}
Once the network is suitably trained, we carry out a calibration test on the quotes of options written on the Microsoft, Costco,  BKNG stocks and the S\&P500 index, respectively (data downloaded from Yahoo Finance).


Table \ref{tab:Descriptive statistics} shows some relevant descriptive statistics of the samples.
The underlying prices of the analyzed options range from 425 USD to 5123 USD, reproducing then a variety of different scales in the data.
We choose a risk-free rate based on the United States treasury bill interest rate up to one-year tenor, which we show in Table \ref{tab:Treasury bill} (this information has been sourced from the U.S.\ treasury department).

\begin{table} 
    \centering
    \caption{
    Descriptive statistics of the traded options in the analyzed equity markets. Underlyings: Microsoft stock, Costco stock, BKNG stock and S\&P500 index.
    }
\begin{tabular}{lccc}
\hline Ticker & Number of samples & Maturity (days) & $\log\left(K / S_0\right)$ \\
\hline MSFT & 192 & {$[38,339]$} & {$[-0.75,0.15]$} \\
 COST & 178 & {$[42,272]$} & {$[-0.72,0.31]$} \\
 BKNG & 204 & {$[37,247]$} & {$[-0.76,0.32]$} \\
 SPX & 212 & {$[40,212]$} & {$[-0.66,0.31]$} \\
\hline
\end{tabular}
\label{tab:Descriptive statistics}
\end{table}

\begin{table} 
    \centering
    \caption{US Treasury bill interest rates (in percentage) as of April 12, 2024.}
    \begin{tabular}{lcccccc}
\hline Weeks & 4 & 8 & 13 & 17 & 26 & 52 \\
 Rate & 5.28 & 5.27 & 5.25 & 5.22 & 5.14 & 4.9 \\
\hline
\end{tabular}
\label{tab:Treasury bill}
\end{table}

In order to fit our DDN to market data we implement a multistart optimization scheme.
That is, we randomly generate multiple combinations of the Heston model parameters as initial values, run a calibration problem for each of these starting points, and select the best solution obtained.
This allows to deal with the possible non-convexity of the objective function.
In addition, as the DDN allows for stable and fast gradient-based optimizations, our multistart scheme is significantly faster than other global optimzation techniques such as stochastic search algorithms.
Specifically, we use the Adam optimizer \citep{kingma2015method} as gradient-based method for the DDN.

As benchmark calibration methods, we consider a calibration based on a feedforward neural network (FNN) that does not embed a differentiation layer, and a Nelder-Mead (N-M) optimization (available from the Python Scipy library, \citealp{virtanen2020scipy}).
For the FNN, we use the same gradient-based optimizer that we use for the DDN.
On the other hand, the N-M is a particularly flexible method that does not require knowledge about the gradient of the objective function.
However, a Nelder-Mead-based calibration is relatively slow and may not be a viable option for a risk manager that deals with frequent changes of the market conditions.
In addition, the N-M depends on the selected initial value, so in order for it to converge to a nearly-global optimum we apply a multistart scheme even in this case.
As the N-M uses the semi-analytical Heston pricing function to calculate the option prices, we do not expect the DDN to produce more accurate calibrations than the N-M.
Instead, we use the N-M results as a benchmark in order to check whether the DDN calibration can enjoy a similar level of accuracy, but in a remarkably lower computational time.

\subsection{Calibration results}\label{sec:calibration_results}

We proceed by showing the calibration results of the DDN, FNN, and N-M methods in terms of the optimized parameter values, the calibration errors, and the computational time.
As error measure we use the mean relative error defined as
\begin{equation}\label{eq:MRE}
    \text { MRE }=\frac{1}{M}\left\{ \sum_{m=1}^M\left|\hat{p}_m-p^{\text{mkt}}_m \right| / p^{\text{mkt}}_m \right\},
\end{equation}
where $p^{\text{mkt}}_m$ denotes the market price of the $m$-th traded option, $\hat{p}_m$ is the corresponding model price computed with the DDN, the FNN, or the semi-analytical pricing function, and $M$ is the number of available market options.



We first show the performance of the three calibration methods under different sizes of the market dataset, namely 10, 50, and 100 traded options in the Microsoft market, respectively.
The results for the N-M, FNN and DDN can be observed in Table \ref{tab:performance of three algorithm}, in which we observe that the N-M algorithm needs a few minutes to reach a sufficiently accurate solution when the market dataset is large.
%
But we immediately notice the little computational time required by the neural network-based methods with respect to the N-M.
Second, we observe that when we consider just 10 market options, the accuracy of the FNN, DDN and N-M are similar in terms of MRE.
However, for larger calibration datasets the FNN exhibits significantly larger errors than the other two methods, while the DDN preserves roughly the same accuracy of the N-M.

\begin{table} 
    \centering
     \caption{
     Mean relative errors (see Eq.\ \eqref{eq:MRE}) and computational times (in brackets) of the Heston model calibration using the Nelder-Mead, the FNN, and the DDN methods, respectively, on 10, 50, and 100 Microsoft call options.
     }
    \begin{tabular}{cccc}
\hline Samples & N-M & FNN & DDN \\
\hline 10 & $0.0064(12.32 \mathrm{s})$ & $0.0068(3.42 \mathrm{s})$ & $0.0067(3.42 \mathrm{s})$ \\ 50 & $0.0175(1 \mathrm{m} 52\mathrm{s})$ & $0.0367(4.12 \mathrm{s})$ & $0.0186(4.13 \mathrm{s})$ \\
 100 & $0.0423(3\mathrm{m} 06\mathrm{s} )$ & $0.0626(7.52 \mathrm{s})$ & $0.0464(7.41 \mathrm{s})$ \\
\hline
\end{tabular}
    \label{tab:performance of three algorithm}
\end{table}

Next, we focus also on the other assets considered in our analysis and we employ the whole sets of traded options (see Table \ref{tab:Descriptive statistics}).
First, we check the variety of volatility dynamics of the assets by performing accurate calibrations and reporting the optimized parameters in Table \ref{tab:Results of parameters calibration}.
The mutually different natures of the analyzed markets remarks the need to use sophisticated models such as the Heston model to describe equity market conditions.

\begin{table} 
\centering
\caption{
Heston parameters calibrated to the four equity markets of Table \ref{tab:Descriptive statistics}.
}
\begin{tabular}{lccccc}
\hline Ticker & $\kappa$ & $\lambda$ & $\sigma$ & $\rho$ & $v_0$ \\
\hline

MSFT &  3.0824 & 0.1477 & 0.7852 & -0.8245 & 0.2514  \\
 COST & 0.0150 & 0.5134 & 0.7775 & -0.5621 & 0.0932  \\
 BKNG & 2.2128 & 0.4093 & 0.8406 & -0.0148 & 0.1102 \\
 SPX &  2.5837 & 0.0269 & 0.4607 & -0.3110 & 0.0567 \\
\hline
\end{tabular}
\label{tab:Results of parameters calibration}
\end{table}

\begin{table} 
\centering
  \caption{
  Heston model calibration results on the equity markets of Table \ref{tab:Descriptive statistics} using the Nelder-Mead, the FNN, and the DDN methods, respectively.
  MRE: mean relative error (see Eq. \eqref{eq:MRE}).
   }
\begin{tabular}{ccccccc}
\cline{2-7}
\multicolumn{1}{l}{} & \multicolumn{2}{c}{N-M}         
& \multicolumn{2}{c}{FNN}       & \multicolumn{2}{c}{DDN}       
\\ \cline{2-7}
& \multicolumn{1}{c}{MRE$^*$}   & Time  & \multicolumn{1}{c}{$|$ MRE$^*-$MRE $|$} & Time & \multicolumn{1}{c}{$|$ MRE$^*-$MRE$|$} & Time \\ \hline
MSFT                   & \multicolumn{1}{c}{0.0604} & 10m12s & \multicolumn{1}{c}{0.0246}     & 7.17s & \multicolumn{1}{c}{0.0013}     & 7.17s \\
COST                   & \multicolumn{1}{c}{0.0554} & 10m16s  & \multicolumn{1}{c}{0.0373}     & 6.45s & \multicolumn{1}{c}{0.0084}     & 6.44s \\
BKNG                   & \multicolumn{1}{c}{0.0642} & 11m43s  & \multicolumn{1}{c}{0.0204}     & 7.34s & \multicolumn{1}{c}{0.0076}     & 7.25s \\
SPX                    & \multicolumn{1}{c}{0.0832} & 12m17s & \multicolumn{1}{c}{0.0169}     & 7.36s & \multicolumn{1}{c}{0.0062}     & 7.36s \\ \hline
\end{tabular}
\label{tab:Comparative Analysis}
\end{table}

Secondly, we compare the calibration performances of the DDN, the FFN, and the N-M across different markets.
We employ the multistart scheme described in Section \ref{sec:calibration_setting}, and note that just a few initial points (about five) are needed in order for the algorithm to reach nearly-global optimum.
%
Despite the calibration errors caused by the selected Heston pricing methodologies (see, e.g., \citealp{levendorskiui2012efficient, levendorskiui2016pitfalls, de2017calibration}), we guarantee as much as possible a fair comparison with the other calibration schemes by using the same initial points of the multistart scheme, and by performing the calibrations mostly on medium-maturity options.

As it is clear from Table \ref{tab:Comparative Analysis},
results are robust to the specific market conditions, and in fact we can draw similar conclusions across different equity products.
As a matter of fact, calibrating the model to the whole datasets of Table \ref{tab:Descriptive statistics} highlights even more the significant computational time required by the Nelder-Mead optimization.
On the other hand, the neural network-based methods converge into a solution in just a few seconds.

\begin{figure} 
    \centering
    \begin{minipage}[b]{0.45\textwidth}
        \centering
        \includegraphics[width=\textwidth]{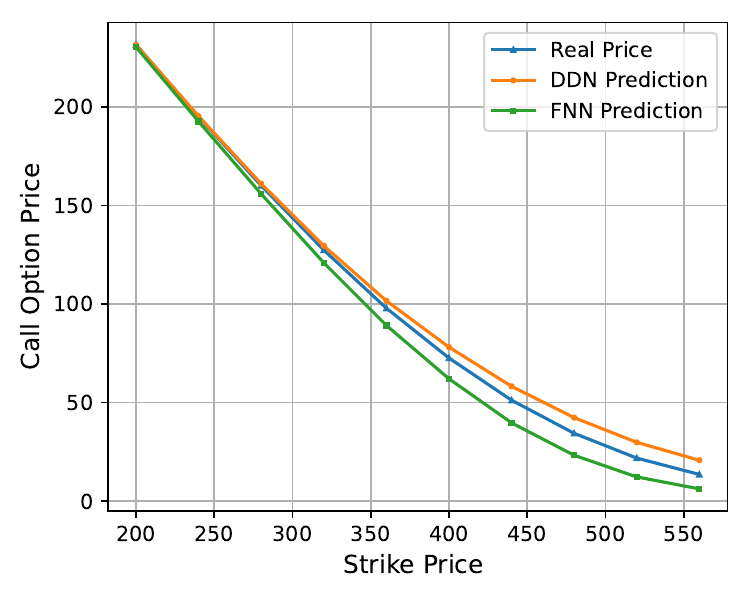}
         \captionsetup{font=small}
        \caption*{(a) Microsoft stock.}
    \end{minipage}
    \hfill
    \begin{minipage}[b]{0.45\textwidth}
        \centering
        \includegraphics[width=\textwidth]{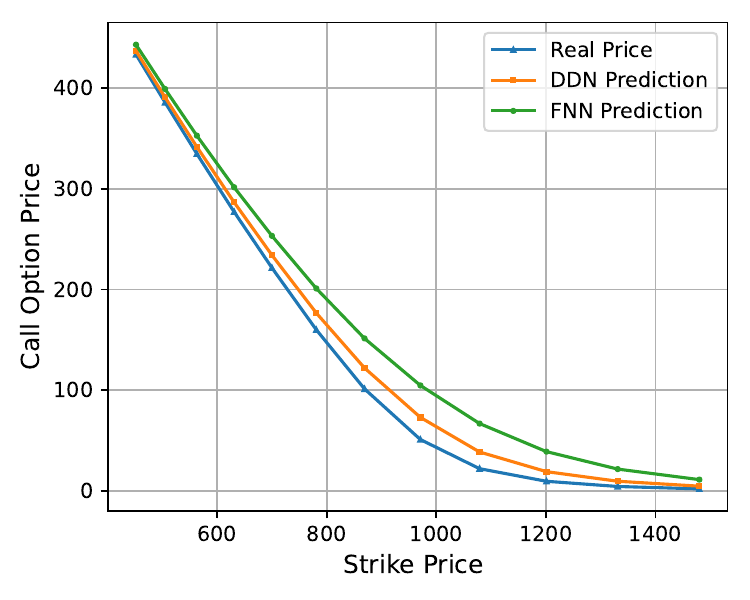}
         \captionsetup{font=small}
        \caption*{(b) Costco stock.}
    \end{minipage}

    \begin{minipage}[b]{0.45\textwidth}
        \centering
        \includegraphics[width=\textwidth]{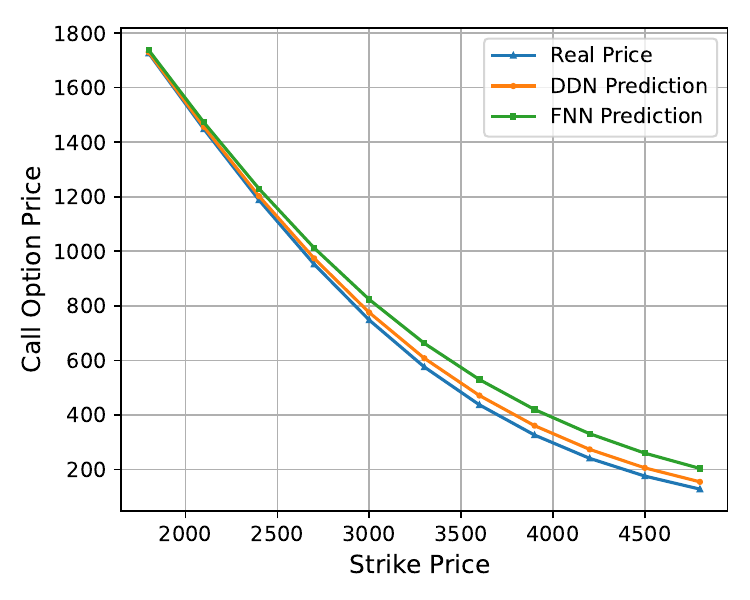}
         \captionsetup{font=small}
        \caption*{(c) BKNG stock.}
    \end{minipage}
    \hfill
    \begin{minipage}[b]{0.45\textwidth}
        \centering
        \includegraphics[width=\textwidth]{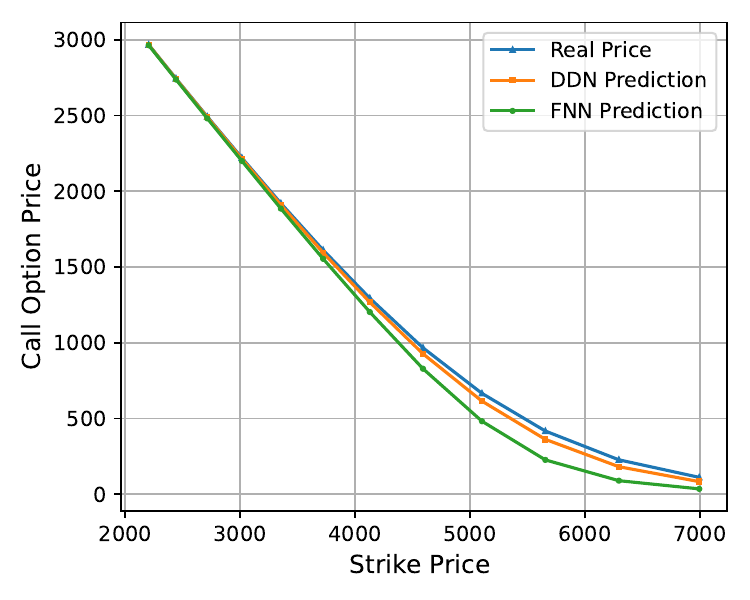}
         \captionsetup{font=small}
        \caption*{(d) S\&P500 index.}
    \end{minipage}
    \caption{
    Market quotes and corresponding DDN and FFN prices of selected call options written on the equity products of Table \ref{tab:Descriptive statistics}, with maturity of 182 days.}
    \label{fig:Real and predicted price}
\end{figure}

We finally provide a visual comparison of the calibration fits of the DDN and the FNN, respectively, in Figure \ref{fig:Real and predicted price}.
We plot the market prices and the model prices of options with selected maturities in the four equity markets.
The DDN prices are significantly close to the market prices as opposed to the FFN prices, confirming the superior performance of our approach.


\section{Conclusions}\label{sec:conclusions}
In this paper we propose a deep differential network (DDN) to learn the plain-vanilla option pricing formula of the Heston stochastic volatility model, and calibrate the model to market data.

Our network estimates both the price and the partial derivatives of the price with respect to the Heston parameters by minimizing an objective function that also includes the partial derivatives of the pricing function.
In this way, our DDN model produces a remarkably good approximation of the Heston function without encountering its numerical issues.
In particular, the DDN finds direct application in the context of model calibration, in which case many evaluation of the pricing formula are needed and the computational speed of the DDN is of crucial importance.
Most importantly, the DDN pricer ensures a stable and reliable computation of its partial derivatives with respect to the model parameters.
As such, the DDN allows to calibrate the model with multistart gradient-based algorithms that significantly outperform the typical global optimizers used in the calibration of option pricing models.

In order to show the validity of our method, we design a number of calibration exercises taking into consideration multiple equity markets.
We show that the DDN produces significantly more accurate calibration results than a feedforward neural network that does not embed a differentiation layer, especially when the calibration dataset is large.
In addition, the DDN achieves roughly the same accuracy of the Nelder-Mead calibration method, which is a gradient-free method widely used in the literature due to its flexibility.
However, the Nelder-Mead converges to the optimum in the order of minutes, while the DDN only requires a few seconds to solve the calibration problem.
Our results are stable to the variety of the assets and market conditions considered in the analysis.

Future researches could implement a differential neural network to calibrate other sophisticated option pricing models and check whether the validity of the DDN persists.
In this regard we highlight jump models, in which estimating sensitivities is complicated and burdensome Monte Carlo simulations may be needed.
For these constructions, it could be particularly convenient to let the network learn the option sensitivities \textit{offline}, so that in the calibration stage we readily dispose of accurate approximations of these partial derivatives.
More in general, it is possible to use the DDN to learn the price sensitivities of any financial derivative.
As such, the DDN can be useful to perform a fast and efficient risk management of exotic products, whose pricing formulas and partial derivatives are typically not analytic.


\noindent \textbf{\\Conflicts of Interest:} The authors declare that there are no conflicts of interest regarding the publication of this article.


\appendix
\clearpage
\newpage
\section{Option prices with QuantLib}\label{app:QuantLib}

\begin{table}[h]
\centering
\caption{
Top four tables: call option prices $\mathcal{G}(\boldsymbol{\theta})$ (see Eq.\ \eqref{eq:heston_oushi}) computed with the \citet{QuantLibReference} pricer for different combinations of inverse log-moneyness $\log(K/S_0)$ and time to maturity $\tau$; specifically, the minimum, medium, and maximum values of the calibration ranges of $\tau$ and $\log(K/S_0)$ as reported in Table \ref{tab:Descriptive statistics} (which shows the global ranges).
Bottom table: sensitivities of the call option prices (computed via numerical differentiation of $\GG$) with respect to the parameters $\boldsymbol{\theta}$ of the stochastic variance process for the medium values of the calibration ranges of $\tau$ and $\log(K/S_0)$.
Maturity in days (annualized in the QuantLib function by multipling by 365).
Underlying initial price $S_0$ set to 1 and continuously compounded risk free $r$ set to 0.
Values of $\boldsymbol{\theta}$ as in Table \ref{tab:Results of parameters calibration}.}
\footnotesize
\begin{tabular}{ccc}
\multicolumn{3}{l}{MSFT}
\\ \hline
$\tau$ & $\log(K/S_0)$ & $\mathcal{G}(\boldsymbol{\theta})$  
\\ \hline
 38 & -0.71 & 0.5084 \\
 38 & -0.28 & 0.2482 \\
 38 & \,\,0.15 & 0.0105 \\
 189 & -0.75 & 0.5314 \\
 189 & -0.31 & 0.2978 \\
 189 & \,\,0.14 & 0.0587 \\
 339 & -0.75 & 0.5373 \\
 339 & -0.31 & 0.3178 \\
 339 & \,\,0.14 & 0.0893 \\
\hline
\end{tabular}
\hspace{10pt}
\begin{tabular}{ccc}
\multicolumn{3}{l}{COST}
\\ \hline
$\tau$ & $\log(K/S_0)$ & $\mathcal{G} (\boldsymbol{\theta})$  
\\ \hline
 42 & -0.69 & 0.4984 \\
 42 & -0.27 & 0.2375 \\
 42 & \,\,0.15 & 0.0017 \\
 157 & -0.72 & 0.5142 \\
 157 & -0.26 & 0.2439 \\
 157 & \,\,0.21 & 0.0074 \\
 272 & -0.69 & 0.5032 \\
 272 & -0.19 & 0.2097 \\
 272 & \,\,0.31 & 0.0070 \\
\hline
\end{tabular}
\\ \vspace{5pt}
\begin{tabular}{ccc}
\multicolumn{3}{l}{BKNG}
\\ \hline
$\tau$ & $\log(K/S_0)$ & $\mathcal{G} (\boldsymbol{\theta})$  
\\ \hline
 37 & -0.68 & 0.4934 \\
 37 & -0.26 & 0.2297 \\
 37 & \,\,0.16 & 0.0056 \\
 142 & -0.61 & 0.4585 \\
 142 & -0.15 & 0.1878 \\
 142 & \,\,0.32 & 0.0220 \\
 247 & -0.76 & 0.5371 \\
 247 & -0.24 & 0.2751 \\
 247 & \,\,0.29 & 0.0657 \\
\hline
\end{tabular}
\hspace{10pt}
\begin{tabular}{ccc}
\multicolumn{3}{l}{SPX}
\\ \hline
$\tau$ & $\log(K/S_0)$ & $\mathcal{G} (\boldsymbol{\theta})$  
\\ \hline
 40 & -0.66 & 0.4831 \\
 40 & -0.25 & 0.2213 \\
 40 & \,\,0.17 & 0.0003 \\
 126 & -0.28 & 0.2459 \\
 126 & \,\,0.02 & 0.0389 \\
 126 & \,\,0.31 & 0.0003 \\
 212 & -0.28 & 0.2485 \\
 212 & \,\,0.02 & 0.0494 \\
 212 & \,\,0.31 & 0.0012 \\
\hline
\end{tabular}
\\ \vspace{10pt}
\begin{tabular}{cccccccc}
\cline{2-8}
& $\tau$ & $\log(K/S_0)$ & $\partial\GG/\partial\kappa$ & $\partial\GG/\partial\lambda$ & $\partial\GG/\partial\sigma$ & $\partial\GG/\partial\rho$ & $\partial\GG/\partial v_0$
\\ \hline
MSFT & 189 & -0.31  & -0.0023 & 0.0768 & \,\,0.0057 & -0.0072  & 0.0820
\\
COST & 157 & -0.26 & \,\,0.0119 & 0.0005 & \,\,0.0050 & -0.0081 & 0.1645
\\
BKNG & 142 & -0.15 & \,\,0.0076 & 0.0718 & -0.0032 & -0.0065 & 0.1430
\\
SPX & 126 & \,\,0.02 & -0.0012 & 0.1977 & -0.0093 & \,\,0.0031 & 0.3667
\\ \hline
\end{tabular}
\label{tab:quantlib}
\end{table}

\end{document}